# Investigating Phosphorus Abundances in a Sample of APOGEE-2 Bulge Globular Clusters

Beatriz Barbuy ,[1] José G. Fernández-Trincado ,[2] Morgan S. Camargo,[1] Doug Geisler ,[3,4]
Maren Brauner ,[5,6] Sandro Villanova ,[7] Dante Minniti ,[8,9] Domingo Anibal García-Hernández ,[5,6]
Stefano O. Souza ,[10] Heitor Ernandes ,[11] Amâncio Friaça ,[1] and Marco Pignatari [12,13,14,15]

[1] Universidade de São Paulo, IAG, Rua do Matão 1226, Cidade Universitária, São Paulo 05508-900, Brazil

[2] Universidad Católica del Norte, Núcleo UCN en Arqueología Galáctica - Inst. de Astrofísica, Av. Angamos 0610, Antofagasta, Chile

[3] Departamento de Astronomia, Casilla 160-C, Universidad de Concepcion, Chile

[4] Departamento de Astronomía, Facultad de Ciencias, Universidad de La Serena. Av. Raul Bitran 1305 , La Serena, Chile

[5] Instituto de Astrofísica de Canarias, C/Via Lactea s/n, E-38205 La Laguna, Tenerife, Spain

[6] Departamento de Astrofísica, Universidad de La Laguna, E-38206 La Laguna, Tenerife, Spain

[7] Universidad Andres Bello, Facultad de Ciencias Exactas, Departamento de Física y Astronomía - Instituto de Astrofísica, Autopista Concepción-Talcahuano 7100, Talcahuano, Chile

[8] Instituto de Astrofísica, Facultad de Ciencias Exactas, Universidad Andres Bello, Fernández Concha 700, Las Condes, Santiago, Chile

[9] Vatican Observatory, Vatican City State 00120, Italy

[10] Max Planck Institute for Astronomy, Königstuhl 17, D-69117 Heidelberg, Germany

[11] Lund Observatory, Department of Geology, Lund University, Sölvegatan 12, Lund, Sweden

[12] Konkoly Observatory, Research Centre for Astronomy and Earth Sciences, HUN-REN, Konkoly Thege Miklós út 15-17, Budapest, H-1121, Hungary

[13] CSFK HUN-REN, MTA Centre of Excellence, Konkoly Thege Miklós út 15-17, Budapest, H-1121, Hungary

[14] University of Bayreuth, BGI, Universitätsstraße 30, 95447 Bayreuth, Germany

[15] NuGrid Collaboration, http://nugridstars.org

## ABSTRACT

Phosphorus-enhanced (P-rich; [P/Fe] $\gtrsim$ +0.8) giants have been found among mildly metal-poor field stars, but in only one star in a globular cluster (GC), M4 (NGC 6121). Also, in a sample of bulge spheroid stars, some of them showed a moderate P-enhancement in the range +0.5 < [P/Fe] < +1.0. In this paper we derive the P abundance of moderately metal-poor ([Fe/H] $\gtrsim$ − 1) GC stars, aiming to check if the phenomenon could be related to the unusual multiple stellar populations found in most GCs. Here we present the detection of P-moderately enhanced stars among two out of seven bulge GCs (Tonantzintla 1, and NGC 6316), with metallicities similar to those of the bulge field P-rich stars. Using $H$-band high-resolution ($R{\sim}22{,}500$) spectra from the APOGEE-2 survey, we present the first high-resolution abundance analysis of [P/Fe] from the PI 16482.932 Å line in a sample of selected bulge GCs. We find that all P-rich stars tend to also be N-rich, that hints at the origin of P-rich stars as second-generation stars in GCs. However no other correlations of P and other elements are found, that are usually indicators of second-generation stars. Further studies with larger samples and comparisons with field stars will be needed before any firm conclusions are drawn.

# 1. INTRODUCTION

Phosphorus-enhanced (P-rich) giant stars were identified by Masseron et al. (2020a) by exceeding a photospheric [P/Fe] $\gtrsim$ + 0.8 (P-rich field stars), abundances well above typical Galactic levels in a range of metallicities,

Corresponding author: Beatriz Barbuy
b.barbuy@iag.usp.br

$-1.35 \lesssim$ [Fe/H] $\lesssim -0.58$, and located in the inner Galactic halo, from the second generation of the Apache Point Observatory Galactic Evolution Experiment (APOGEE-2; Majewski et al. 2017). The sample was further investigated by Brauner et al. (2023) and Brauner et al. (2024), including a detailed discussion of the possible nucleosynthesis processes responsible for the enhancement of *phosphorus*.

Recently, we analyzed [P/Fe] abundances in a sample of 58 stars, selected from the reduced proper motion (RPM) stars from Queiroz et al. (2021), having kinematical and dynamical characteristics indicating that they belong to the bulge spheroid, and with [Fe/H]$\sim$−1.0 (Razera et al. 2022). For these stars, we studied the abundances of C, N, O, Mg, Si, Ca and Ce in Razera et al. (2022), Na and Al in Barbuy et al. (2023), and the iron-peak elements (V, Cr, Mn, Co, Ni and Cu) in Barbuy et al. (2024). In a fourth paper, Barbuy et al. (2025) derived abundance ratios of phosphorus, sulfur, and potassium for this sample. In particular, [P/Fe] abundances were not available from the latest APOGEE-2 data release (see e.g., Abdurro'uf et al. 2022). We found that our sample has similarities with that of Masseron et al. (2020a), Brauner et al. (2023), and Brauner et al. (2024), although less P-enhanced, with P abundances reaching [P/Fe]$\sim$ +1.0. As discussed by Barbuy et al. (2025), such a variations in the P abundances between different studies are due to the different methods and stellar parameters adopted. Despite these uncertainties, the existence of P-enhanced stars is a puzzle for stellar nucleosynthesis to be explained.

The Galactic bulge is the least studied region of the Galaxy (e.g. Barbuy et al. 2018), and much is still to be revealed from observations of bulge field (e.g. Nepal et al. 2025) and globular cluster (GC) stars (e.g. Souza et al. 2024a). In the present work we search for P-rich stars among selected bulge GCs that also have metallicity around the P-rich field stars, and that were analyzed in the *H*-band, based on high-resolution ($R \sim$22,500) near-infrared (NIR) spectra from the APOGEE-2 survey (Majewski et al. 2017; Schiavon et al. 2024), and the bulge Cluster APOgee Survey subprogram (CAPOS; Geisler et al. 2021).

The inspected clusters included the APOGEE-2 clusters NGC 6522 (Fernández-Trincado et al. 2019), UKS 1 (Fernández-Trincado et al. 2020), Tonantzintla 1 or NGC 6380 (Fernández-Trincado et al. 2021a, hereafter Ton 1), and the CAPOS clusters Tonantzintla 2 or Pismis 26 (Fernández-Trincado et al. 2022, hereafter Ton 2), NGC 6558 (González-Díaz et al. 2023), HP 1 Henao et al. (2025) and NGC 6316 (Frelijj et al. 2025), with some stars in NGC 6380 also observed within CAPOS. Our aim is to verify whether the excess on [P/Fe] found in moderately metal-poor field stars is also detected in bulge GCs in the same metallicity range.

This work is organized as follows: In Section 2, we describe the sample stars. In Section 3, we describe the spectrum synthesis calculations and results, and their implications are discussed in Section 4. In Section 5 our conclusions are drawn.

## 2. THE SAMPLE

In the present work, we use data from the seventeenth data release (DR 17; Abdurro'uf et al. 2022) of the APOGEE-2 survey (Majewski et al. 2017), which is one of the programs of the Sloan Digital Sky Survey IV (Blanton et al. 2017), to investigate the [P/Fe] abundance ratios for selected bulge GCs, which are not available from APOGEE-2 DR 17 data products. The APOGEE-2 survey collects high-resolution ($R \sim 22,500$) spectra with high signal-to-noise ratios in the *H*-band (15140–16940 Å) from NIR wavelength spectrographs (Wilson et al. 2019) that operated on the Sloan 2.5 m telescope (Gunn et al. 2006) at the Apache Point Observatory (Northern Hemisphere) and on the Irénée du Pont 2.5 m telescope (Bowen & Vaughan 1973) at Las Campanas Observatory (Southern Hemisphere). The targeting strategy of the APOGEE-2 survey is summarized in Beaton et al. (2021) and Santana et al. (2021), while the spectra are reduced as described in Nidever et al. (2015), and analyzed using the APOGEE Stellar Parameters and Chemical Abundance Pipeline (ASPCAP; García Pérez et al. 2016). The analysis of *H*-band spectra in the APOGEE-2 project is carried out through a Nelder-Mead algorithm (Nelder & Mead 1965), which simultaneously fits the stellar parameters —effective temperature ($T_{eff}$), gravity (log g), metallicity ([Fe/H]), and microturbulence velocity ($v_t$), together with the abundances of $\alpha$-elements, and C, N, and O abundances, with the ASPCAP pipeline, which is based on the FERRE code (Allende Prieto et al. 2006). We revised the critical [C/Fe], [N/Fe], [O/Fe] ASPCAP abundances, given the presence of blending molecular lines on the PI lines.

The sample consists mainly of giant star members of the GCs Ton 1, and NGC 6316, for which a P-enrichment has been detected in this study.

For HP-1 there might be 1 to 3 stars with P-excess, but a low S/N prevents a firm conclusion. We also inspected the [P/Fe] abundance ratios in the clusters NGC 6522, UKS 1, NGC 6558, and Ton 2, however, no P-rich stars were identified for these GCs.

## 3. PHOSPHORUS ABUNDANCES

We computed the phosphorus abundances from the PI 15711.622 and 16482.932 Å lines in the *H*-band, adopting the oscillator strengths log $gf = -0.510$ and $= -0.273$ respectively, from the APOGEE-2 collaboration. The PI 15711.622 Å line, although computed for all stars, is shallow, so we concentrate our analysis on the PI 16482.932 Å line. In any case, the shallower PI 15711.622 Å line fit was either compatible with the result of the better line, or too shallow to be considered. These are the only useful lines in the *H*-band, and the same lines used by Masseron et al. (2020a); Brauner et al. (2023). We used the code for spectrum synthesis TURBOSPECTRUM from Alvarez & Plez (1998) and Plez (2012), together with atmospheric models from Gustafsson et al. (2008). The solar abundance adopted for P is that of Asplund et al. (2021), that is, A(P) = 5.41, and those of the $^{12}C^{14}N^{16}O$ trio, here very important due to the blending $^{12}C^{16}O$ lines, A(C)=8.46, A(N)=7.83, and A(O)=8.69. The complete atomic line list used is that from the APOGEE collaboration, together with the molecular lines described in Smith et al. (2021).

In Table 1 we report the photometric and ASPCAP spectroscopic (uncalibrated) stellar parameters, including effective temperature ($T_{eff}$), surface gravity (log $g$), metallicity [Fe/H], and microturbulence velocity ($v_t$) for stars of the globular clusters Ton 1 and NGC 6316. These stars were analyzed, from the APOGEE-2 spectra, by Fernández-Trincado et al. (2021a), and Frelijj et al. (2025), respectively, and we adopt their stellar parameters. For Ton 1 in particular Fernández-Trincado et al. (2021a) derived a series of photometric and spectroscopic parameters. Although the study carried out by Fernández-Trincado et al. (2021a) was based on spectra from the APOGEE-2 DR 16 (Ahumada et al. 2020) with low-S/N ratios, this was certainly improved with a second observation visit which was available in the APOGEE-2 DR 17, for which spectra with high-S/N (S/N>115 pixel$^{-1}$) for almost all the Ton 1 members were made available. Therefore, we used the latter spectra for our analysis. We adopted their photometric parameters, whereas their spectroscopic ones are very close to the ASPCAP DR17 parameters, which we also adopted, for a double check. The abundances of [C/Fe], [N/Fe], and [O/Fe] were revised, for internal consistency, because of their importance for properly taking into account the blending of molecular CO lines. For this we used the region $15525 - 15590$ Å, which contains lines of $^{12}C^{16}O$, $^{16}OH$, and $^{12}C^{14}N$, as described in Barbuy et al. (2021) and Razera et al. (2022). The strength of CO lines, that affect the main PI line, are further checked with the line CO 15717.2 Å near the secondary PI line, as illustrated in the Appendix A. These revised CNO abundances are reported in Table 1.

We also analyzed stars from NGC 6522, NGC 6558, UKS 1, and Ton 2, with stellar parameters of Fernández-Trincado et al. (2019), González-Díaz et al. (2023), Fernández-Trincado et al. (2020), Fernández-Trincado et al. (2021b), and Fernández-Trincado et al. (2022), respectively. We found that these stars do not show any P-enhancements, a relevant result that will be discussed below. Finally, HP 1 analysed by Henao et al. (2025) does show a P line that could be enhanced in 3 out of 10 stars, but the lower S/N prevents us from making a conclusion, so that we count this cluster as not exhibiting P-enhancement.

We also calculated the abundances of C, N, O and P, using the uncalibrated, or spectroscopic stellar parameters, from the ASPCAP procedure, available in APOGEE-2 DR 17. As concluded by da Silva et al. (2024), the ASPCAP method (García Pérez et al. 2016), which used the balance of molecular lines of CO, CN and OH, is a reliable indicator of stellar parameters, in particular effective temperatures. This double work was worth the effort to check the reliability of high values of N in an important fraction of the sample stars. Interestingly, several stars for which the P line is detected, show a very high N abundance, typical of second generation stars of globular clusters. This double check was also useful because, for Ton 1 we identified 7 out of 12 stars with high levels of [P/Fe] $\gtrsim +0.5$. For NGC 6316 we have identified 6 P-rich stars ([P/Fe] $\gtrsim +0.5$). In contrast, Frelijj et al. (2025) reported [P/Fe] abundances for only four stars in this cluster, whose values seem different by about $\pm0.3$ dex from ours, perhaps due to a difference in the continuum placement, a difference in the atmospheric models, or to the noise. Uncertainties are reported in Table A.3.

Figure 1 shows examples of the fitted PI 16482.932 Å lines for four stars in the clusters Ton 1 (*rows 1* and *2*) and NGC 6316 (*rows 3* and *4*).

## 4. DISCUSSION

In this Section we discuss the possibility of P being produced together with N in second-generation stars, and compare our results with literature data and chemical evolution models.

### 4.1. *Nitrogen vs. Phosphorus*

The high N abundance in many of the P-rich stars is striking. First, we stress that the high abundances of N are compatible between all sources, i.e. the abundances from CAPOS and APOGEE-2 photometric and spectroscopic parameters, as well as from our derivations of C, N and O, using all of the above parameters, as can be seen in Table 1.

**Table 1.** Globular clusters (GCs) with high [P/Fe] in our sample. Column information: (1) GC ID; (2) APOGEE star identification; (3–6) stellar parameters taken from literature (*first line*) and ASPCAP (*second line*); columns (7), (8), (9), (10) report our measured [C/Fe], [N/Fe], [O/Fe], and [P/Fe] abundance ratios. In column (11) are reported the S/N from DR 17.

| GC | APOGEE-ID | $T_{eff}$ (K) | log $g$ (cgs) | [Fe/H] | $v_t$ (km s$^{-1}$) | [C/Fe] | [N/Fe] | [O/Fe] | [P/Fe] | S/N pixel$^{-1}$ |
|---|---|---|---|---|---|---|---|---|---|---|
| (1) | (2) | (3) | (4) | (5) | (6) | (7) | (8) | (9) | (19) | (11) |
| Ton 1 | 2M17342588−3901406 | 3584 | 0.26 | −0.78 | 2.37 | +0.20 | +0.30 | +0.60 | <0.40 | 260 |
|  |  | 3650 | 0.64 | −0.87 | 2.44 | +0.40 | +0.30 | +0.75 | <0.40 |  |
| Ton 1 | 2M17341922−3906052 | 3866 | 0.77 | −0.61 | 1.56 | +0.25 | +0.35 | +1.00 | <0.40 | 162 |
|  |  | 3808 | 1.01 | −0.80 | 2.17 | +0.40 | +0.25 | +0.65 | <0.40 |  |
| Ton 1 | 2M17342921−3904514 | 3850 | 0.74 | −0.79 | 2.50 | −0.20 | +1.00 | +0.40 | +0.60 | 242 |
|  |  | 4047 | 1.17 | −0.74 | 2.00 | +0.00 | +1.20 | +0.90 | +0.25 |  |
| Ton 1 | 2M17343616−3903344 | 3890 | 0.81 | −0.85 | 2.31 | +0.10 | +1.10 | +0.65 | +0.50 | 142 |
|  |  | 3982 | 1.15 | −0.79 | 2.11 | +0.15 | +1.80 | +1.10 | +1.00 |  |
| Ton 1 | 2M17343025−3903190 | 3859 | 0.75 | −0.85 | 2.48 | −0.14 | +1.02 | +0.35 | +0.70 | 145 |
|  |  | 4117 | 1.21 | −0.73 | 2.17 | +0.15 | +0.80 | +0.70 | +0.50 |  |
| Ton 1 | 2M17342693−3904060 | 4073 | 1.14 | −0.90 | 2.21 | +0.20 | +0.85 | +0.60 | <0.40 | 160 |
|  |  | 4152 | 1.29 | −0.84 | 2.18 | +0.15 | +0.45 | +0.40 | <0.40 |  |
| Ton 1 | 2M17342541−3902338 | 4059 | 1.12 | −0.86 | 2.27 | +0.30 | +0.63 | +0.40 | +0.40 | 88 |
|  |  | 4282 | 1.57 | −0.72 | 2.00 | +0.25 | +0.55 | +0.50 | +0.80 |  |
| Ton 1 | 2M17342767−3903405 | 4125 | 1.24 | −0.79 | 2.10 | −0.10 | +1.10 | +0.45 | +1.00 | 144 |
|  |  | 4307 | 1.62 | −0.68 | 1.93 | −0.10 | +0.80 | +0.35 | +0.80 |  |
| Ton 1 | 2M17341969−3905457 | 4099 | 1.19 | −0.83 | 2.14 | −0.20 | +1.20 | +0.35 | +0.80 | 146 |
|  |  | 4357 | 1.67 | −0.69 | 1.99 | +0.00 | +0.70 | +0.35 | +0.50 |  |
| Ton 1 | 2M17342177−3906173 | 4166 | 1.31 | −0.78 | 2.10 | −0.20 | +1.20 | +0.40 | +0.70 | 123 |
|  |  | 4345 | 1.73 | −0.69 | 1.94 | −0.15 | +1.00 | +0.40 | +0.50 |  |
| Ton 1 | 2M17342943−3902500 | 4136 | 1.26 | −0.76 | 2.01 | +0.20 | +0.50 | +0.00 | +0.70 | 125 |
|  |  | 4362 | 1.68 | −0.69 | 1.98 | +0.20 | +0.60 | +0.20 | +0.80 |  |
| Ton 1 | 2M17343521−3903091 | 4174 | 1.33 | −0.87 | 2.11 | −0.40 | +1.00 | +0.20 | +0.80 | 115 |
|  |  | 4607 | 1.57 | −0.73 | 2.66 | +0.30 | +0.65 | +0.40 | +0.40 |  |
| NGC 6316 | 2M17163864−2809385 | 4119 | 1.18 | −0.88 | 2.49 | +0.40 | +0.80 | +1.00 | +0.80 | 65 |
|  |  | 4003 | 1.43 | −0.73 | 1.73 | +0.20 | +0.40 | +0.60 | +0.80 |  |
| NGC 6316 | 2M17165235−2809502 | 4118 | 1.18 | −0.74 | 1.67 | +0.25 | +0.45 | +0.95 | <0.40 | 70 |
|  |  | 3996 | 1.39 | −0.75 | 1.71 | +0.28 | +0.55 | +0.70 | <0.40 |  |
| NGC 6316 | 2M17163623−2808067 | 3842 | 0.68 | −0.86 | 2.95 | −0.20 | +0.95 | +0.30 | +0.40 | 80 |
|  |  | 4059 | 1.21 | −0.77 | 2.16 | +0.00 | +0.90 | +0.60 | +0.40 |  |
| NGC 6316 | 2M17163330−2808396 | 3979 | 0.93 | −0.84 | 2.24 | −0.40 | +1.55 | +0.45 | +0.80 | 85 |
|  |  | 4094 | 1.35 | −0.73 | 2.08 | +0.15 | +1.60 | +0.70 | +0.70 |  |
| NGC 6316 | 2M17164048−2808443 | 4051 | 1.06 | −0.85 | 2.17 | +0.10 | +0.90 | +0.65 | +0.40 | 85 |
|  |  | 4071 | 1.52 | −0.73 | 1.73 | +0.15 | +0.65 | +0.55 | +0.40 |  |
| NGC 6316 | 2M17164482−2808302 | 4083 | 1.12 | −0.76 | 1.69 | +0.20 | +0.55 | +0.85 | +0.70 | 90 |
|  |  | 4004 | 1.36 | −0.76 | 1.72 | +0.20 | +0.35 | +0.65 | +0.80 |  |
| NGC 6316 | 2M17163393−2811052 | 3884 | 0.75 | −1.00 | 2.70 | +0.15 | +0.90 | +0.70 | +0.40 | 105 |
|  |  | 3932 | 1.25 | −0.81 | 1.86 | +0.15 | +0.80 | +0.60 | +0.40 |  |
| NGC 6316 | 2M17163903−2807212 | 3871 | 0.73 | −0.97 | 2.80 | +0.15 | +0.95 | +0.75 | +0.75 | 135 |
|  |  | 3871 | 0.94 | −0.76 | 2.00 | +0.12 | +0.85 | +0.70 | +0.50 |  |
| NGC 6316 | 2M17163911−2804506 | 3754 | 0.52 | −0.85 | 2.23 | +0.30 | +0.42 | +0.85 | <0.40 | 140 |
|  |  | 3710 | 0.58 | −0.90 | 2.45 | +0.32 | +0.20 | +0.70 | <0.40 |  |
| NGC 6316 | 2M17163627−2807166 | 3668 | 0.36 | −0.93 | 2.72 | +0.20 | +0.55 | +0.55 | +0.30 | 170 |
|  |  | 3666 | 0.73 | −0.82 | 2.47 | +0.17 | +0.30 | +0.45 | <0.40 |  |

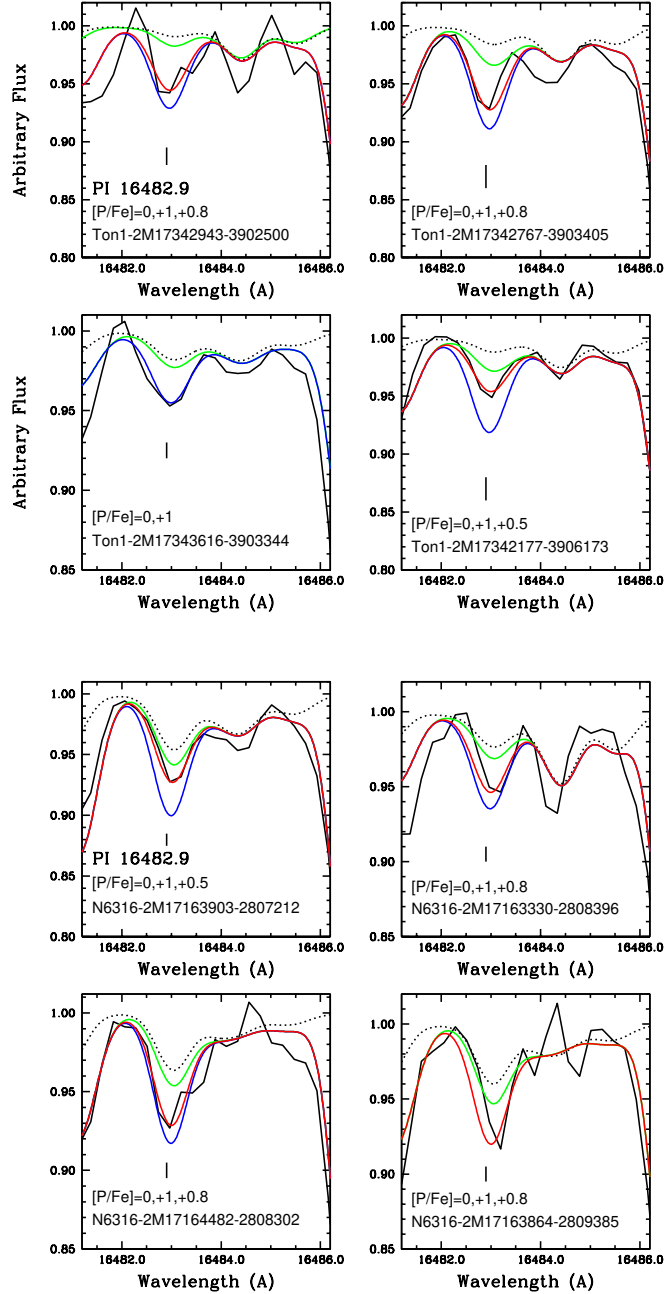

**Figure 1.** P I 16482.932 Å line in 4 stars in Ton 1 (*rows 1 and 2*) and NGC 6316 (*rows 3 and 4*), fitted with synthetic spectra computed with [P/Fe]= 0.0 (green), 1.0 (blue, and final values (red) if different from 0.0 or 1.0. Synthetic spectra computed with molecular lines only are shown as dotted lines.

It appears that the P-rich stars tend to be also N-rich, but the contrary is not true, i.e., not all N-rich stars are P-rich even in the same GC. A connection P:N could be a hint on the possibility that the P-enhancement phenomenon occurs in second-generation stars of globular clusters. However, there is no clear one-to-one correlation between the P and the N abundances, given that in these cases the P abundance tends to be [P/Fe]=0.8, although there is a correlation with N/O. Besides, Brauner et al. (2023) found that not all P-rich stars are also N-rich, although many of them are N-rich as well. It is unclear if this difference between our present results and field P-rich stars is real, or it is just due to the small stellar sample currently available from GCs. Nevertheless, our stellar data in Figure 2 show a clear correlation between [P/Fe] versus [N/O], in particular for stars with the higher N abundances in Ton 1 and NGC 6316. In the same figure we also inspected Al abundances: Ton 1 exhibits a Al-rich-N-rich behaviour similar to that observed in atypical Al-rich-N-rich field stars Fernández-Trincado et al. (2020); but this is less clear in NGC 6316 (possibly due to selection effects). In any case, there is evidence that N-Al-rich stars originate in GCs (Fernández-Trincado et al. 2021b; Minniti et al. 2024; Souza et al. 2024b).

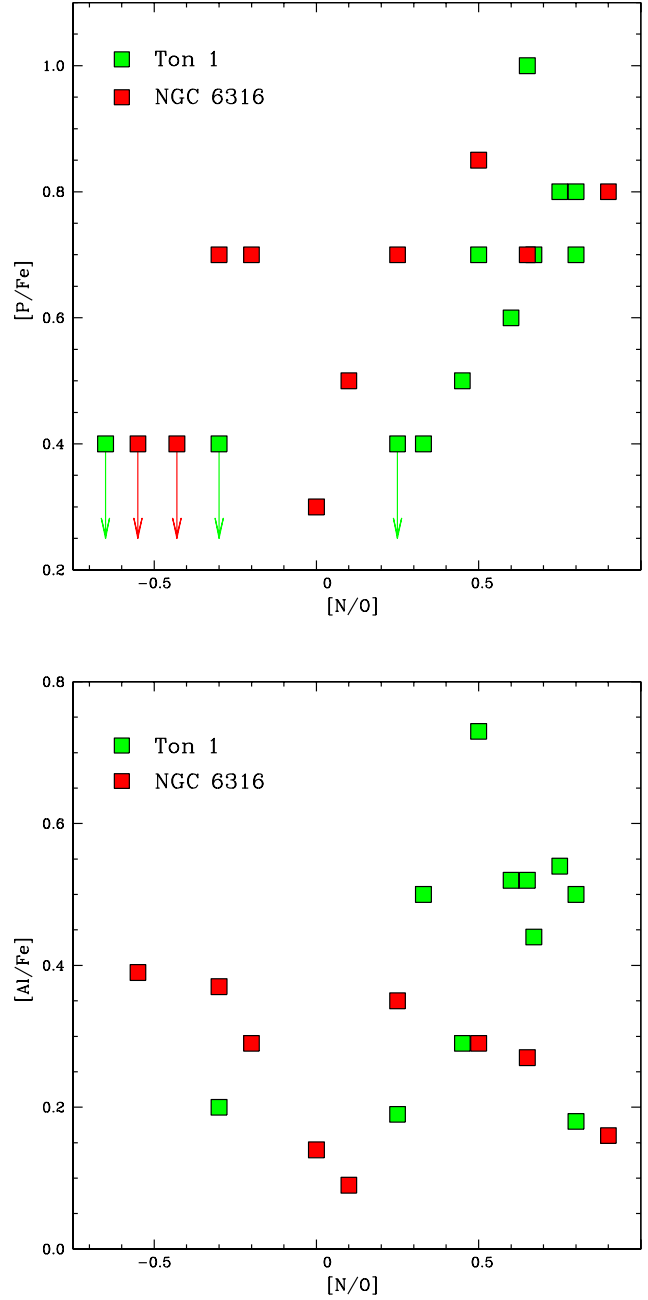

**Figure 2.** [P/Fe] vs. [N/O] (upper panel) and [Al/Fe] vs. [N/O] (lower panel) for Ton 1 (green squares) and NGC 6316 (red squares).

## 4.2. *Chemical evolution models*

Figure 3 shows the present results compared with data from the literature and chemical evolution models from Barbuy et al. (2025). Data from the literature on P, compared to other elements, are relatively rare. Previous work includes Caffau et al. (2011) in 20 disk dwarf stars, and Caffau et al. (2016) added another 4 disk dwarf stars. Roederer et al. (2014) measured P in 14 halo stars. Maas et al. (2019) derived P abundances in 9 disk stars, and Maas et al. (2022) in 163 disk and halo stars. Sadakane & Nishimura (2022) studied P in 45 main-sequence stars, Nandakumar et al. (2022) measured P in 38 disk stars.

Finally, Masseron et al. (2020a,b) measured P in 16 stars, Brauner et al. (2023) enlarged the sample to 78 confirmed P-rich stars (including the ones from Masseron et al. (2020a,b)). An important result of Brauner et al. (2023) was the detection of a member enriched in P of the globular cluster M4, a nearby disk cluster. Brauner et al. (2024) further explored in detail the anomalous pattern of heavy elements for a small sample of P-rich stars, resulting in enhanced heavy element abundances, possibly of s-process and/or i-process origin (Lugaro et al. 2023). These very P-rich stars are not included in Figure 3, in order to not hide the effect on the clusters.

Masseron et al. (2020a), Brauner et al. (2023), and Brauner et al. (2024) have obtained very high P abundances ([P/Fe] > +0.8) for a sample of stars that have metallicities around [Fe/H]∼−1. The orbital analysis of the Masseron-Brauner sample of P-rich stars led to the identification of most of them belonging to the thick disk and the inner Galactic halo. However, no such effect is seen at higher or lower metallicities. In particular, the results by Roederer et al. (2014) for halo stars show that such an effect is absent in the outer halo.

In Brauner et al. (2024) the nucleosynthesis processes responsible for P-enhancement were discussed, but while CO-shell mergers from massive stars (Ritter et al. 2018; Andrassy et al. 2020; Roberti et al. 2025)] were the most promising sources for the abundance anomalies of P-rich stars, the astrophysical source of P-rich stars is not yet established.

Barbuy et al. (2025) have performed chemical evolution modeling in order to investigate the abundance pattern of phosphorus in bulge spheroid stars. Our chemical evolution models for the bulge spheroid include not only hypernovae but also enhancement of odd-Z elements from the neutrino-process taking place during core-collapse supernova yields (CCSN; Yoshida et al. 2008). However, these nucleosynthetic ingredients are effective only at very low metallicities and still leave unexplained the P-excess of mildly metal-poor stars. Our chemical evolution models (Barbuy et al. 2025) predict a maximum [P/Fe]= +0.45 at [Fe/H] ∼−0.85.

## 4.3. *Bulge stellar populations*

We analyzed here the [P/Fe] abundances in the bulge globular clusters Ton 1, and NGC 6316. Note that, interestingly, these GCs are in the main APOGEE-2 SDSS-IV and also CAPOS sample. The same applies to HP 1, for which a detection has still to be confirmed. However, so are NGC 6522, NGC 6558, UKS 1, and Ton 2, which are not P-rich. We would like to identify a similarity between these two clusters that show P-enriched stars, in contrast to those that do not show P-enrichment. They could indeed characterize a subsample of clusters, or is this an effect of self-enrichment, and indicate some type of particular nucleosynthesis. In Table 2 we list their main characteristics. We do not see significant differences between the two samples of clusters.

Massari et al. (2019) suggested Ton 1 and NGC 6316, as well as HP 1, NGC 6522, and NGC 6558, to be in-situ main bulge clusters, and Ton 2 a low-energy cluster, whereas UKS 1 was not classified. Pérez-Villegas et al. (2020) classified HP 1, Ton 1, NGC 6522, NGC 6558, as in situ main bulge clusters, and Ton 2, and NGC 6316 as thick disk members. Forbes (2020) suggested Ton 2 to belong to Koala. Callingham et al. (2022) proposed HP 1, Ton 1, NGC 6522, and NGC 6558 as main bulge, and Ton 2 and NGC 6316 as members of their suggested `Kraken` structure. Belokurov & Kravtsov (2024) suggested Ton 1, Ton 2, HP 1, NGC 6316, NGC 6522, NGC 6558, and UKS 1 to be in-situ main bulge clusters. Geisler et al. (2025) classifies all our samples as bulge GCs except Ton 2 (Disk) and UKS 1, which is uncertain as to whether it is in-situ or ex-situ. The P-rich phenomenon appears to be present in both bulge and disk GCs. We also note that Ton 1 and Ton 2 are closely projected in the sky with a separation of ∼40', but their distances show that they do not form a physical pair.

In conclusion, we find no clear distinction in any major parameter between the P-rich and P-normal GCs, except perhaps their current masses, with a trend for P-rich stars to be in clusters with masses above $10^5$ M$_\odot$, but the currently most massive GC, NGC 6522, does not show P enhancement. It should be noted that the samples are small, and the clusters with the largest samples are the ones that show P-enhanced stars.

## 5. CONCLUSIONS

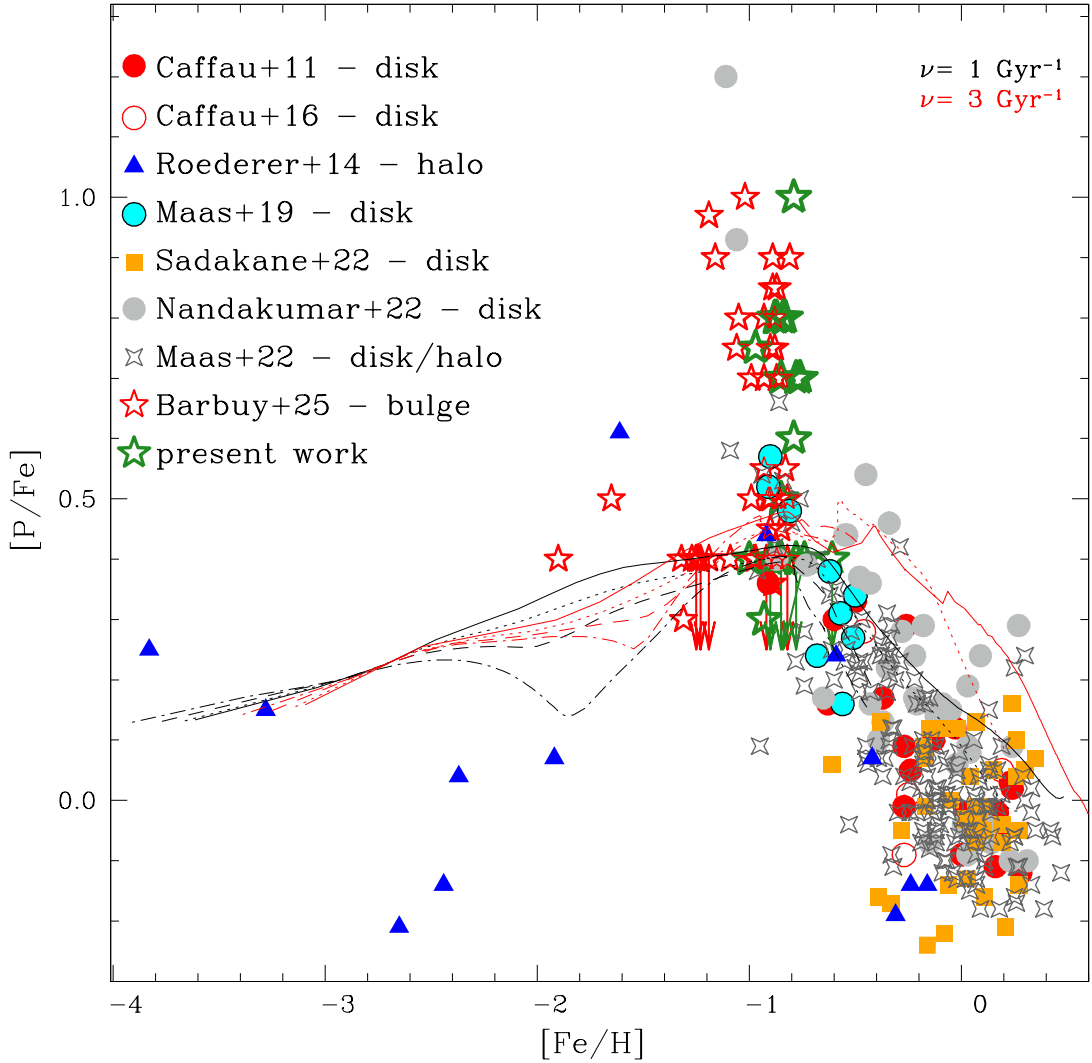

**Figure 3.** [P/Fe] vs. [Fe/H] for the present results compared with literature data. Symbols – darkgreen-open stars: present work, red-filled circles: Caffau et al. (2011), red- open circles: Caffau et al. (2016), blue-filled triangles: Roederer et al. (2014), filled-cyan circles + open-black circles: Maas et al. (2019), light grey-filled circles: Nandakumar et al. (2022), light grey open 4-side stars: Maas et al. (2022), red-open stars: Barbuy et al. (2025). Different model lines correspond to the outputs of models computed for radii r < 0.5, 0.5 < r < 1, 1 < r < 2, and 2 < r < 3 kpc from the Galactic centre. Black lines correspond to specific star formation $\nu = 1$ Gyr$^{-1}$, red lines to $\nu = 3$ Gyr$^{-1}$.

We computed [P/Fe] abundance ratios in bulge GCs, in order to verify which of them could host P-rich stars. In five of them, namely NGC 6522, NGC 6558, UKS 1, and Ton 2, (and possibly HP 1), no star showed P-enhancement, On the other hand in Ton 1, and NGC 6316, more than half of the analyzed stars do show P-enhancements, in the range of +0.5 < [P/Fe] < +1.0. These values are not as high as those found by Masseron et al. (2020a), Brauner et al. (2023), and Brauner et al. (2024) in field thick disk and inner-halo stars, but they are of the same order as those found in the sample of 58 bulge field stars by Barbuy et al. (2025), with kinematical, dynamical, and chemical characteristics of the early bulge spheroid. All of these stars have metallicities between [Fe/H]≈ −1.2 and −0.7.

From our results, it appears that all P-rich stars tend to be also N-rich, and that only a fraction of the analysed stars show this effect, in the same metallicity range as the field stars. This could be the consequence of the so-called *second-generation* phenomenon that affect GCs. However, there is no clear correlation between the abundances of P and N, and we found no correlation between P and other indicators as second-generation as Mg. For Ton 1 there is a N:Al correlation, but not for NGC 6316.

**Table 2.** List of analysed globular clusters. Masses and radial velocities from Baumgardt & Hilker (2018), Fernández-Trincado et al. (2020), distances from Baumgardt & Vasiliev (2021). Metallicities are taken from Fernández-Trincado et al. (2021a) for Ton 1, Fernández-Trincado et al. (2022) for Ton 2, Henao et al. (2025) for HP 1, Frelijj et al. (2025) for NGC 6316, Fernández-Trincado et al. (2019) for NGC 6522, González-Díaz et al. (2023) for NGC 6558, and Fernández-Trincado et al. (2020) for UKS 1; and ages from the compilation by Deras et al. (2023) and Bica et al. (2024). $N_{stars}$:x is the number of stars studied, and x is the number of P-rich stars with [P/Fe]>0.5.

| Cluster | Mass ($M_\odot$) | RV (kms.$^{-1}$) | $d_\odot$ (kpc) | [Fe/H] | Age (Gyr) | $N_{stars}$ |
|---------|------|-----|------|--------|-----|---------|
| Ton 1 | $3.41 \times 10^5$ | $+1.92 \pm 2.4$ | $9.607 \pm 0.30$ | $-0.80 \pm 0.04$ | $12.9 \pm 1.1$ | 12:7 |
| NGC 6316 | $3.75 \pm 0.69 \times 10^5$ | $+99.1 \pm 0.8$ | $11.152 \pm 0.38$ | $-0.87 \pm 0.02$ | $13.1 \pm 0.5$ | 10:6 |
| HP 1 | $1.1 \pm 0.38 \times 10^5$ | $+40.1 \pm 1.1$ | $6.995 \pm 0.14$ | $-1.15 \pm 0.03$ | $12.8 \pm 0.9$ | 10:1 |
| NGC 6522 | $3.92 \pm 0.54 \times 10^5$ | $-14.0 \pm 0.6$ | $7.295 \pm 0.21$ | $-1.04 \pm 0.09$ | $12.8 \pm 1.0$ | 5:0 |
| NGC 6558 | $2.93 \pm 1.09 \times 10^4$ | $-195.6 \pm 0.7$ | $7.474 \pm 0.29$ | $-1.15 \pm 0.08$ | $12.3 \pm 1.1$ | 4:0 |
| UKS 1 | $8.0 \times 10^4$ | $+66.12 \pm 12.9$ | $15.581 \pm 0.57$ | $-0.98 \pm 0.11$ | $13.0 \pm 1.2$ | 6:0 |
| Ton 2 | $8.01 \pm 4.02 \times 10^4$ | $-183.8 \pm 0.8$ | $6.987 \pm 0.34$ | $-0.70 \pm 0.05$ | 12 | 7:0 |

In summary, our results strengthen the connection between moderately metal-poor GCs and bulge field stars, because we detect the presence of P-rich stars in two bulge GCs. P-rich stars might be second generation stars (as evidenced by the simultaneous presence of P-rich and P-normal stars in a single GC, and that all P-rich stars tend to be N-rich), which would align with the massive star nucleosynthesis scenario proposed here. On the other hand, the strongly P-rich field stars by Brauner et al. (2023) show a range of N abundances, making the production of P-rich stars more complex than the second-generation GC explanation. Therefore this scenario should be further investigated in the future. It would indeed be an elegant solution for the puzzling metallicity distribution of P-rich field stars, to identify them as stars escaped [MP: are they escaping, accreted from disrupted GCs, or else?] from old bulge GCs. This would explain why there are no P-rich stars at solar-like metallicity or at much lower metallicities typical of the MW halo.

On the other hand, certain nucleosynthesis scenarios, such as those involving novae (Smith & Kraft 1996; Iliadis et al. 2016), can be discarded to explain P-rich stars, due to the finding of P-rich stars in two old GCs, where there was no time for novae to evolve and contribute to stellar chemical enrichment, because the time delay to form novae is similar to that to form supernovae type Ia, a time of about ∼1 Gyr (Cescutti & Molaro 2019).

If we consider the main parameters of the GCs studied here, there is no significant difference between those showing P-excess, and the others. The P-overabundance was not observed in other stellar populations, although P is not often studied in the literature. Further investigations are needed on the P-excess, through the analysis of other samples of stars and from the nucleosynthesis side. Specifically, the observed excess P would exclude an AGB scenario, since AGB stars do not produce P efficiently (e.g. Karakas & Lattanzio 2014; Pignatari et al. 2016). Finally, if the connection P-rich stars and second stellar generations in GC is confirmed, that would hint at massive stars as main polluters in GCs, at least with respect to P.

Another important result is the absence of P-rich stars in five of the seven clusters studied, and the key question that remains is: Why we can find P-enriched stars in some clusters and not in other similar clusters? Higher signal-to-noise ratios, and larger samples of stars in the detected P-enhanced clusters would be of great interest. A quantitative theoretical explanation of the origin of P-excess in bulge GCs (and in P-rich field stars) is still missing, and more work is needed. As we mentioned, the number of stars with [P/Fe] is still limited, and more in general it is still matter or debate what are the effective stellar polluter(s) producing the abundance pattern of SG stars in GCs. P enrichment and the correlations found with other elements could at least help to clarify their occurrence in different GCs.

We are grateful to T. Masseron for helpful comments. B.B. and A.C.S.F. acknowledge grants from FAPESP, Conselho Nacional de Desenvolvimento Científico e Tecnológico (CNPq) and Coordenação de Aperfeiçoamento de Pessoal de Nível Superior (CAPES) - Financial code 001. J.G.F-T gratefully acknowledges the grants support provided by ANID Fondecyt Postdoc No. 3230001 (Sponsoring researcher), and from the Joint Committee ESO-Government of Chile under the agreement 2023 ORP 062/2023. M.S.C. acknowledges a CAPES doctoral fellowship. D.G. also acknowledges financial support from the Dirección de Investigación y Desarrollo de la Universidad de La Serena through the Programa de Incentivo a la Investigación de Académicos (PIA-DIDULS). D.G. gratefully acknowledges the support provided by Fondecyt regular n. 1220264. D.M. gratefully acknowledges support from the Center for Astrophysics and Associated Technologies (CATA) by ANID BASAL projects ACE210002 and FB210003, and Fondecyt Project No. 1220724. M.B. acknowledges financial support from the European Union and the State Agency of Investigation of the Spanish Ministry of Science and Innovation (MICINN) under the grant PRE-2020-095531 of the Severo Ochoa Program for the Training of Pre-Doc Researchers (FPI-SO). S.V. gratefully acknowledges the support provided by Fondecyt Regular n. 1220264 and by the ANID BASAL project FB210003. S.O.S. acknowledges the support from Dr. Nadine Neumayer's Lise Meitner grant from the Max Planck Society. H.E. acknowledges a post-doctoral fellowship at Lund Observatory. M.P. acknowledges the support to NuGrid from the "Lendulet-2023" Program of the Hungarian Academy of Sciences (LP2023-10, Hungary), the ERC Synergy Grant Programme (Geoastronomy, grant agreement number 101166936, Germany), the ERC Consolidator Grant funding scheme (Project RADIOSTAR, G.A. n. 724560, Hungary), the ChETEC COST Action (CA16117), supported by the European Cooperation in Science and Technology, and the IReNA network supported by NSF AccelNet (Grant No. OISE-1927130). MP also thanks the support from NKFI via K-project 138031 (Hungary). We acknowledges support from the ChETEC-INFRA project funded by the European Union's Horizon 2020 Research and Innovation programme (Grant Agreement No 101008324). This work benefited from interactions and workshops co-organized by The Center for Nuclear astrophysics Across Messengers (CeNAM) which is supported by the U.S. Department of Energy, Office of Science, Office of Nuclear Physics, under Award Number DE-SC0023128.

Funding for the Sloan Digital Sky Survey IV has been provided by the Alfred P. Sloan Foundation, the U.S. Department of Energy Office of Science, and the Participating Institutions. SDSS-IV acknowledges support and resources from the Center for High-Performance Computing at the University of Utah. The SDSS website is www.sdss.org.

## APPENDIX

### A. UNCERTAINTIES

There are no published NLTE corrections for the P I lines analyzed in this work, therefore these uncertainties cannot be evaluated here. Figure 4 gives [P/Fe] vs. the stellar parameters effective temperature and gravity, showing that there is essentially no trend between these quantities.

A typical uncertainty is computed by adopting errors in the stellar parameters of $\Delta T_{eff} = 100$ K, $\Delta \log g = 0.2$, $\Delta v_t = -0.2$ km s$^{-1}$, shown in Table 3. This is applied to the cool star Ton 1: 2M17342588−3901406 with $T_{eff}$ of 3584 K. The total error in [P/Fe] is of 0.1 dex. For this star the CO is strong enough within their low temperature range such that a variation of 100K does not change much the derived P abundance. In contrast, the example of a less cool star Ton1: 2M17343616-3903344, which shows a larger difference in P abundance between the literature and the ASPCAP parameters with $T_{eff} = 3890$ K and 3982 K, respectively – in this particular case, there is a weakening of the CO lines, leading to a higher abundance of P with the warmer parameters (ASPCAP). The fits are shown in Figure 5. For stars in this temperature range the uncertainty can be considered to reach up to 0.3 dex, due to a change in the CO band intensity. It is important to stress that the higher P abundances are found for stars with $T_{eff} > 4000$ K, therefore much less affected by the CO line intensity blending the PI line.

Finally, the use of a unique line, with only in some cases the weaker line being useful, is a vulnerability of the present results. It would be of interest to observe other P lines in these same targets - although in the optical the extinction in the bulge in much more problematic.

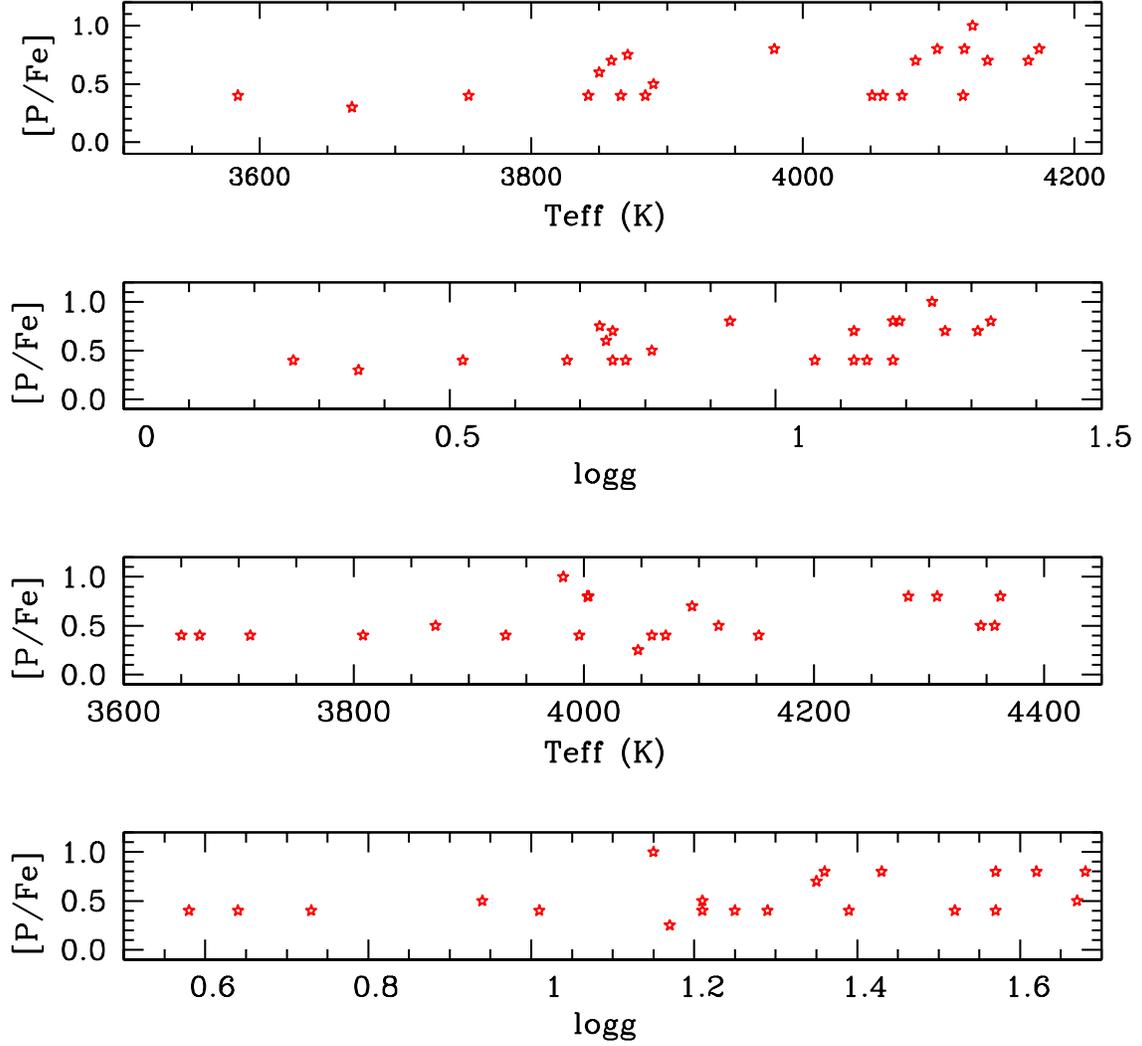

**Figure 4.** [P/Fe] vs. $T_{eff}$ and [P/Fe] vs. log g, for stellar parameters from references cited in the text, and ASPCAP parameters.

**Table 3.** Phosphorus abundance uncertainties for star 1: Ton 1: 2M17342588-3901406, and star 2: Ton 1: 2M173436163903344. We compute the uncertainties for changes in the stellar parameters of $\Delta T_{eff} = 100$ K, $\Delta$log g = 0.2, $\Delta v_t = -0.2$ km s$^{-1}$ and the corresponding total error is given in the last column.

| Star | [P/Fe] | $\Delta T$ $\pm 100$ K | $\Delta$log $g$ $\pm 0.2$ dex | $\Delta v_t$ $\pm 0.2$ km.s$^{-1}$ | $(\sum x^2)^{1/2}$ |
|------|--------|------|------|------|------|
| Star 1 | +1.0 | −0.01 | +0.1 | +0.0 | +0.1 |
| Star 2 | +0.5 | +0.2 | +0.05 | +0.15 | +0.25 |

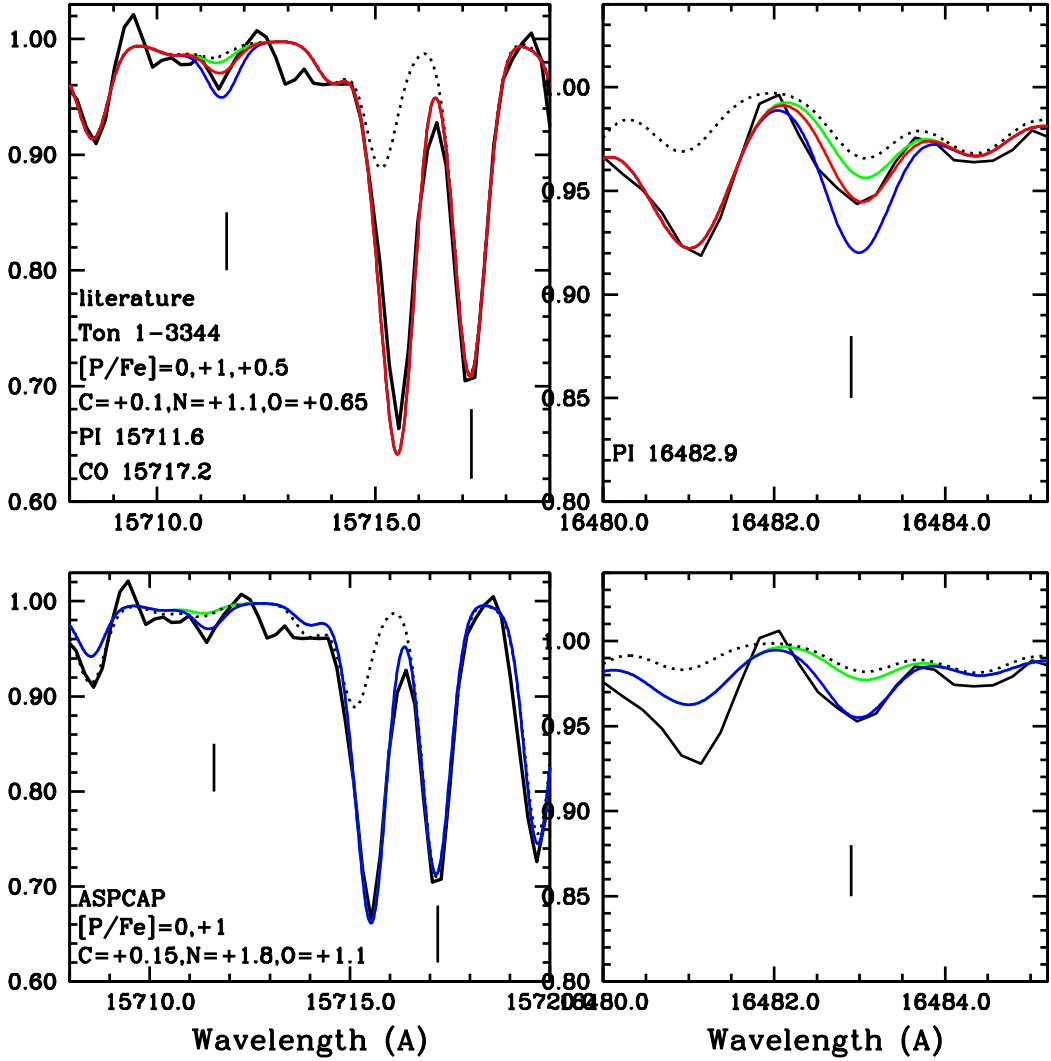

**Figure 5.** PI lines and CO lines in star Ton1: 2M17343616-3903344 for stellar parameters from Fernández-Trincado et al. (2021c) and ASPCAP parameters, with $T_{eff}$ = 3890 K and 3982 K, respectively. CO lines intensity changes more dramatically in the temperature range.